\documentclass[twocolumn]{aastex631}

\usepackage{amsmath}
\usepackage{enumitem}
\usepackage{multirow}
\usepackage{hyperref}
\usepackage{graphicx}

\received{}
\revised{}
\accepted{}
\submitjournal{}

\shorttitle{SPIRITS19q}

\begin{document}

\title{SPIRITS 19q: Dust Production by a Subsolar-metallicity Carbon-rich Wolf-Rayet Star in NGC 2403}

\author[0000-0003-4725-4481]{Sam Rose} 
\correspondingauthor{Sam Rose}
\email{srose@caltech.edu}
\affiliation{Division of Physics, Mathematics, and Astronomy, California Institute of Technology, Pasadena, CA 91125, USA}

\author[0000-0003-0778-0321]{Ryan M.\ Lau}
\affil{IPAC, Mailcode 100-22, Caltech, 1200 E.\ California Blvd., Pasadena, CA 91125, USA}
\author[0000-0001-5754-4007]{Jacob E.\ Jencson}
\affil{IPAC, Mailcode 100-22, Caltech, 1200 E.\ California Blvd., Pasadena, CA 91125, USA}

\author[0000-0002-5619-4938]{Mansi M.\ Kasliwal}
\affil{Division of Physics, Mathematics, and Astronomy, California Institute of Technology, Pasadena, CA 91125, USA}

\author[0000-0002-2090-9751]{Andreas A.C. Sander}
\affil{Zentrum für Astronomie der Universität Heidelberg, Astronomisches
Rechen-Institut, Mönchhofstr. 12-14, 69120 Heidelberg, Germany}
\affil{Institut f{\"u}r Theoretische Physik und Astrophysik, Christian-Albrechts-Universit{\"a}t zu Kiel, Leibnizstr.\ 15, 24118 Kiel, Germany}

\author[0000-0003-1377-7145]{Howard E. Bond}
\affil{Department of Astronomy and Astrophysics, Penn State University, University Park, PA 16802, USA}
\affil{Space Telescope Science Institute, 3700 San Martin Dr., Baltimore, MD 21218, USA}

\author{Michael F. Corcoran}
\affil{CRESST II and X-ray Astrophysics Laboratory NASA/GSFC, Greenbelt, MD 20771, USA}
\affil{Institute for Astrophysics and Computational Sciences, The Catholic University of America, 620 Michigan Ave., N.E. Washington, DC 20064, USA}

\author[0000-0002-9129-5988]{Izumi Endo}
\affil{Department of Astronomy, Graduate School of Science, The University of Tokyo, Bunkyo-ku, Tokyo 113-0033, Japan}

\author[0000-0001-9315-8437]{Matthew  J.\ Hankins}
\affil{Arkansas Tech University, 215 West O Street, Russellville, AR 72801, USA}

\author[0000-0003-4870-5547]{Olivia C.\ Jones}
\affil{UK Astronomy Technology Centre, Royal Observatory, Blackford Hill, Edinburgh, EH9 3HJ, UK}

\author[0000-0003-2758-159X]{Viraj R. Karambelkar}
\affil{ Columbia University, 538 West 120th Street 704, MC 5255, New York, NY 10027, USA}

\author{Astrid Lamberts}
\affiliation{Universit\'e C\^ote d’Azur, Observatoire de la C\^ote d’Azur, CNRS, Laboratoire Lagrange, Bd de l’Observatoire, CS 34229, F-06304 Nice cedex 4, France}

\author[0000-0001-7697-2955]{Thomas Madura}
\affil{Department of Physics and Astronomy, San Jose State University, San Jose, CA 95192-0106, USA}

\author{Anthony F.J. Moffat}
\affiliation{D\'epartement de physique, Universit\'e de Montr\'eal, 1375 avenue
Th\'er\`ese-Lavoie-Roux, Montr\'eal, QC, H2V 0B3, Canada}

\author{Patrick W.\ Morris}
\affiliation{California Institute of Technology, Pasadena, CA 91125, USA}

\author[0000-0002-8234-6747]{Takashi Onaka}
\affil{Department of Astronomy, Graduate School of Science, The University of Tokyo, Bunkyo-ku, Tokyo 113-0033, Japan}

\author[0000-0001-5644-8830]{Michael E.\ Ressler}
\affil{Jet Propulsion Laboratory, California Institute of Technology, MS 169-327, 4800 Oak Grove Drive, Pasadena, CA 91109, USA}

\author[0000-0002-2806-9339]{Noel D. Richardson}
\affiliation{Department of Physics and Astronomy, Embry-Riddle Aeronautical University, 3700 Willow Creek Rd, Prescott, AZ 86301, USA}

\author{Christopher M. P. Russell}
\affil{Department of Physics and Astronomy, Bartol Research Institute, University of Delaware, Newark, DE 19716 USA}

\author[0000-0001-7641-5497]{Itsuki Sakon}
\affiliation{Department of Astronomy, Graduate School of Science, The University of Tokyo, Bunkyo-ku, Tokyo 113-0033, Japan}

\author{J. Sanchez-Bermudez}
\affiliation{Universidad Nacional Aut\'onoma de M\'exico, Instituto de Astronom\'ia,  Apdo. Postal 70264, Ciudad de M\'exico, 04510, M\'exico}

\author[0000-0001-9754-2233]{Gerd Weigelt}
\affil{Max-Planck-Institut f\"ur Radioastronomie, Auf dem H\"ugel 69, 53121 Bonn, Germany}

\author[0000-0002-8092-980X]{Peredur M.~Williams}
\affil{Institute for Astronomy, University of Edinburgh, Royal Observatory, Edinburgh EH9 3HJ, UK}

\begin{abstract}


We present JWST/NIRSpec IFU observations of SPIRITS~19q, the highly dust-producing carbon-rich (WC) binary candidate located in a subsolar-metallicity region of the nearby spiral galaxy NGC~2403. The observations, taken in April of 2024, confirm the association of a dusty outburst observed in 2019 by the \textit{Spitzer Space Telescope\/} with an early-type WC star. Using models from the Potsdam Wolf-Rayet (PoWR) LMC model grid we find that the WC star of SPIRITS 19q likely has an especially high mass-loss rate ($\gtrsim$ 10$^{-4}$ $M_{\odot}$ yr$^{-1}$). From the flux peak of the IR transient as measured by \textit{Spitzer}/IRAC as well as constraints on dust composition and size from the JWST spectrum, we estimate a total dust mass formed in the outburst of 6.6 $\pm$ 0.4 $\times$ 10$^{-6}$ $M_{\odot}$. Assuming a minimum orbital period of 12 years, this corresponds to a period-averaged dust production rate of $\lesssim$ 5.5 $\times$ 10$^{-7}$  $M_{\odot}$ yr$^{-1}$. These observations suggest that even a single WC system can contribute to the dust budget at subsolar metallicities, and that such systems are an important source of carbonaceous dust grains in the early universe. 

\end{abstract}

\section{Introduction} 
\label{sec:Introduction}

In the last decade there has been growing recognition of the ubiquitous presence of dust across cosmic time and in a variety of environments, even while our understanding of how that dust is formed has remained incomplete. 
It is unclear if known channels of dust formation including the winds of low-mass (0.8--8 $M_{\odot}$) evolved Asymptotic Giant Branch (AGB) stars, as well as supernovae explosions (SNe), which can produce dust in the expanding ejecta at late times (100--1000 days post-explosion), can fully account for the amount of dust observed in the early universe, and at low metallicities in the local universe \citep{Dwek:2011, Boyer:2012, Lesniewska:2019, Viero:2022, Witstok:2023}.
In order for AGB stars to contribute significantly to dust production, a sufficient amount of time since star formation must elapse before the lower-mass stars have evolved off the main sequence. 
SNe, which are produced by higher-mass stars ($>$ 8 $M_{\odot}$) with much shorter lifetimes, can produce dust earlier, but it is unknown how much dust survives after the explosion \citep[e.g.,][]{ Micelotta:2016A&A...590A..65M,Priestley:2021MNRAS.500.2543P, Martinez:2022, Schneider:2024A&ARv..32....2S}.

Another potentially important source of dust in the early universe, as well as at low metallicities in the local universe, could be dust production by carbon-rich Wolf-Rayet (WC) binary systems, as suggested in \citet{Lau:2020}.
Wolf-Rayet (WR) stars are a hot (T$_{*}$ $\gtrsim$ 25,000 K) and luminous (L$_{*}$ $\gtrsim$ 10$^5$ $L_{\odot}$) evolved stage of massive stars characterized by broad emission lines from fast-moving (v $\gtrsim$ 500-1,000 km s$^{-1}$), dense, helium-rich, stellar winds \citep{Crowther:2007}. 
Classical WR stars are classified based on the occurrence and ratio of
prominent spectral emission lines. The three main types are labelled
after the prominent occurrence of nitrogen (WN), carbon (WC), and oxygen
(WO) in the spectrum. The WN surface composition is a product of the CNO
cycle, while the WC and the  rarer WO stars show enhanced surface
abundances of carbon and oxygen, implying that these objects are further
stripped than WN stars. The ratio of emission lines with different
ionization stages defines subtypes within each of the three main types,
effectively yielding to the situation that later subtypes have cooler
photospheres.

Dust-production is observed in WC stars with OB type companions in the dense self-shielded regions of the shock-fronts produced by the colliding winds of the WC star and its companion \citep[e.g.,][]{Allen:1972A&A....20..333A, Williams:1987QJRAS..28..248W, Usov:1991}. The dust-formation is often bursty, matching the orbital period of the binary as dust is formed around each periastron passage while the WC star and its companion are near their minimum orbital separations. 

In our own galaxy, WR 140 represents a particularly dramatic example with a highly eccentric orbit with a period of 7.9 years which leads to periodic dust-formation episodes every orbital period \citep{Williams:1990, Usov:1991, Williams:2009}. 
Recent JWST/MIRI images from \citet{Lau:2022} and  \citet{Lieb:2025ApJ...979L...3L} resolve the spiral rings of dust expanding outward associated with each outburst. 
Incorporating dust production into the Binary Population and Spectral Synthesis (BPASS) code of \citet{Eldridge:2017}, \citet{Lau:2020} showed that these WC systems can produce significant amounts of dust at sub-solar metallicity. 

Part of the reason that WC-OB dust-production has been overlooked in favor of AGB stars or SNe is that such systems were expected to be rare, as WR stars were thought to be formed only from the most massive progenitors (M$_{*}$ $\gtrsim$ 40 $M_{\odot}$). 
Alternate pathways for WC formation through binary stripping and interaction \citep[e.g.,][]{Paczynski:1967AcA....17..355P} are available.
With an increasing understanding of the prevalence of binarity for most massive stars \citep{Sana:2012, Smith:2014}, the importance of this formation channel has been reevaluated.
Recent spectroscopic results from \citet{Dsilva:2020} suggest that the fraction of WC stars in the Milky Way with orbiting companions is at least 70$\%$. There are two confirmed examples of dust-forming WC+O systems at sub-solar metallicity in the LMC, HD~36402 and HD~38030  \citep{Williams:2013, Williams:2021}, which may have been produced through the companion stripping mechanism. 
Dust formed around WC stars may also not be destroyed as some massive stars will not explode as supernovae \citep{Gerke:2015MNRAS.450.3289G, Burrows:2025ApJ...987..164B, De:2026Sci...391..689D, Maltsev:2025A&A...700A..20M}.

To date there have been no confirmed highly dust-producing WC-OB binaries at lower metallicity than the LMC, which could be used to test predictions made by theoretical models \citep{Lau:2020}. 
In this work we present JWST NIRSpec observations of one such potential dust-producing WC system at sub-solar metallicity. 

SPIRITS 19q is one of six candidate extragalactic dust-producing WC systems presented in \citet{Lau:2021}. 
It was initially identified based on mid-IR variability detected by the SPitzer InfraRed Intensive Transients Survey (SPIRITS) survey \citep{Kasliwal:2017, Karambelkar:2019}. 
SPIRITS was an IR transient survey of 194 nearby galaxies (within 20 Mpc) using the Spitzer Space Telescope \citep{Werner:2004, Gehrz:2007}, which ran from 2014 until Spitzer was decommissioned in 2020 January.  
SPIRITS used the InfraRed Array Camera (IRAC) \citep{Fazio:2004} in Channel~1 ($3.6\,\mu$m) and Channel~2 (4.5 $\mu$m) to obtain observations with a baseline cadence between one week and six months, achieving a 5$\sigma$ depth of 20.0~mag and 19.1~mag (Vega system), in Channels~1 and~2, respectively.  

SPIRITS 19q lies in the outskirts of the nearby galaxy NGC~2403 \citep[located 3.2 Mpc from the Milky Way;][]{Radburn-Smith:2011}, at J2000 coordinates 07:37:18.21 +65:33:49.0.
An increase in mid-IR emission from SPIRITS 19q was first observed by Spitzer on 2017 June 26 (MJD 57930). 
The emission peaked on 2018 June 20, and has been fading ever since at a rate of $\sim$ 0.4 mag yr$^{-1}$ at 3.6 $\mu$m and $\sim$ 0.3 mag yr$^{-1}$ at 4.5 $\mu$m.
At peak the mid-IR emission was $-14.49$ $\pm$ 0.01 mag at 3.6 $\mu$m and $-15.12$ $\pm$ 0.01  at 4.5 $\mu$m \citep[in absolute Vega magnitudes][]{Lau:2021}.

\begin{figure*}
    \centering
    \includegraphics[scale=0.7]{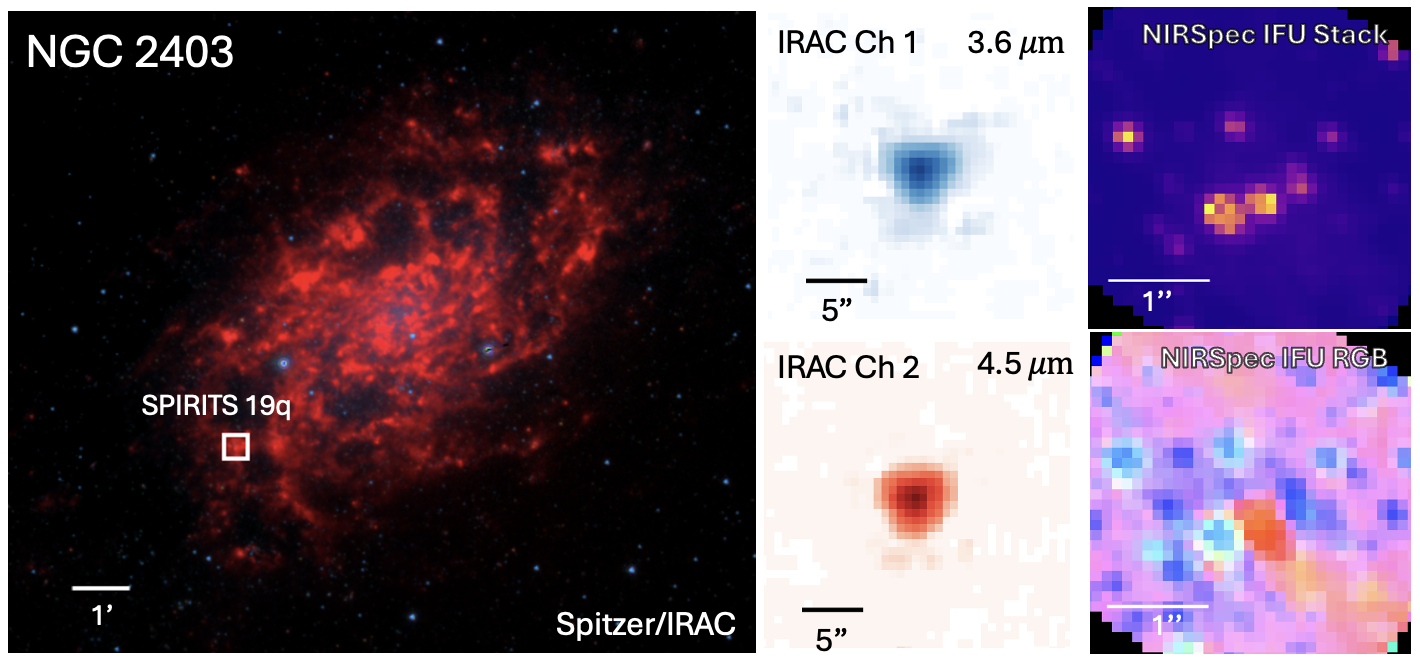}
    \caption{\textit{Left:} False-color (blue Channel 1 (3.6 $\mu$m), green Channel 2 (4.5 $\mu$m), and red Channel 4 (7.9 $\mu$m)) Spitzer/IRAC image of NGC 2403 showing the location of SPIRITS 19q. \textit{Top Center:} Spitzer/IRAC Channel 1 image of SPIRITS 19q with reference image subtraction from \citet{Lau:2021}. \textit{Bottom Center:} Spitzer/IRAC Channel 2 image of SPIRITS 19q with reference image subtraction from \citet{Lau:2021} \textit{Top Right:} JWST/NIRSpec IFU image of SPIRITS 19q summed over the full PRISM band-pass 0.6--5.3 $\mu$m. \textit{Bottom Right:} False color JWST/NIRSpec IFU image (blue 0.6--2.2 $\mu$m, green 2.2--3.7 $\mu$m, and red 3.7-5.3 $\mu$m). At the distance of NGC 2403 the 0.1" spaxels of the NIRSpec IFU correspond to a region $\sim$ 1.55 pc$^{2}$ in size. We identify SPIRITS 19q as the reddest object in the 3"x3" IFU field of view and note that its position is consistent with the measured centroid location from \textit{Spitzer}. }
    \label{fig:spirits19q_inset_fig}
\end{figure*}

In quiescence, SPIRITS 19q has relatively bright mid-IR emission ($<-12.5$ at 3.6 $\mu$m and $-13.0$ at 4.5 $\mu$m in absolute Vega magnitudes), suggesting that it is located within an \ion{H}{2} region \citep{Lau:2021}. 
Optical Keck/LRIS spectra also presented in \citet{Lau:2021} have narrow line emission which further supports this conclusion.
Its location in NGC 2403 further indicates that it is in the VS 51 \citep{Veron:1965} \ion{H}{2} region which lies a deprojected distance of 3.5 kpc from the core of the galaxy \citep{Garnett:1997}.
The location of SPIRITS 19q within NGC 2403 is shown in Figure~\ref{fig:spirits19q_inset_fig}.
\citet{Moustakas:2010} measure the 12 + log(O/H) oxygen abundance of VS 51 to be 8.43 $\pm$ 0.13 which is less than the solar value of 8.69 $\pm$ 0.05 \citep{Asplund:2009} implying that SPIRITS 19q is located in a subsolar metallicity region (Z$_{\mathrm{VS 51}}$ = 0.6 Z$_{\odot}$).

In addition to resembling a WC colliding-wind binary, the SPIRITS 19q outburst observed by \textit{Spitzer} could also be explained by a massive stellar merger \citep{Smith:2016} or an intermediate-luminosity red transient \citep[ILRT, see for example;][]{Bond:2009}, however they both should be rarer than a WC dust-producing binary system and thus less likely.
The optical and near-IR spectra presented in \citet{Lau:2021} include spectral features suggesting that the bright \ion{H}{2} region which hosts SPIRITS 19q also hosts an early type WC star.
It is necessary to confirm the association of the WC star with the dusty outburst to positively rule out other candidate outburst explanations which, because SPIRITS 19q is within a crowded region of stars in VS 51, requires the high spatial resolution of JWST. 

In this work we use JWST/NIRSpec IFU observations to confirm the first subsolar-metallicity highly dust producing Wolf-Rayet system.
In section \ref{sec:Observations} we discuss the new observations of SPIRITS 19q which allow us to test the association of the WC star with the dusty outburst. 
In section \ref{sec:Results and Analysis} we analyze these observations to conclude that the WC star is associated with the dusty outburst. 
In section \ref{sec:Discussion} we consider the implications for dust-formation in subsolar-metallicity environments and at high redshifts based on our results. 
Finally, in section \ref{sec:Conclusions} we finish with a summary of our main conclusions. 

\section{Observations} 
\label{sec:Observations}

\subsection{\textit{Spitzer}/IRAC}

In Figure \ref{fig:lightcurve} we show the mid-IR photometry for SPIRITS 19q obtained by \textit{Spitzer}/IRAC \citet{Lau:2021} and WISE/NEOWISE (section \ref{sec:wise_photo}) as well as the mid-IR color evolution. As noted in \citep{Lau:2021} mid-IR color distinguishes dust continuum from the free-free emission, originating in the dense ionized wind, which dominates the MIR emission of dust-free WC stars. In the case of SPIRITS 19q, the red colors during the outburst indicate substantial dust formation. 

\subsection{Keck I/LRIS}
\label{lris}

\citet{Lau:2021} presented optical spectroscopy of SPIRITS 19q obtained using the Low Resolution Imaging
Spectrometer on Keck I on UT 2019-04-03, 287 days after the peak of the MIR outburst observed by \textit{Spitzer}. Here we present an additional LRIS spectrum taken on UT 2026-02-19 to observe SPIRITS 19q in quiescence. We obtained raw data from the Keck Observing Archive (KOA) for the observations presented in \citet{Lau:2021}. Both the old and the new observation were taken with the D560 dichroic, 400/8500 red grating, the 400/3400 blue grism, and the long 1.0'' slit. We reprocessed both sets of observations with the \texttt{lpipe} \citep{lpipe} pipeline. The observed spectra are very similar (Figure \ref{fig:LRIS}), suggesting that there is very little contamination of the wind emission features from emission in the shock interaction region in the first epoch.

\subsection{JWST/NIRSpec Observations} 
\label{JWST/NIRSpec Observations}

The JWST observations of SPIRITS 19q were taken under Cycle 1 GO program (PI Lau, ID 1863) on 2024 April 4 (MJD=60404).
A NIRSpec \citep{Jakobsen:2022A&A...661A..80J} IFU observation was obtained in the PRISM/Clear disperser-filter configuration to obtain spatially resolved observations (0.1'' spaxels which, at the distance of NGC 2403, corresponds to a region $\sim$ 1.55 pc$^2$, about the size of a compact star cluster) of the bright \ion{H}{2} region in which SPIRITS 19q was identified. This observation covered 0.6--5.3 $\mu$m with an average spectral resolving power of $\sim$100. The relatively broad wavelength coverage as well as the unmatched spatial resolution of \textit{JWST} allow us to unequivocally associate the WC features observed in our previous ground-based observations with the red continuum of the dusty outburst observed by \textit{Spitzer}. 


Raw (\texttt{uncal}) files were downloaded from the Mikulski Archive for Space Telescopes (MAST)\footnote{\url{https://mast.stsci.edu/portal/Mashup/Clients/Mast/Portal.html}} and reduced using a local installation of the standard JWST Science Calibration Pipeline\footnote{\url{https://jwst-pipeline.readthedocs.io/en/stable/jwst/introduction.html}} version 1.20.2 under the Calibration Reference Data System (CRDS; \citealp{Greenfield:2016A&C....16...41G}) context defined by \texttt{jwst\_1464.pmap}.

\begin{deluxetable}{ccc}
\tablecaption{NEOWISE MIR Photometry of SPIRITS 19q
    \label{tab:photometry}}
\tablehead{
  \colhead{\hspace{0.5cm}MJD}\hspace{0.5cm} & 
  \colhead{\hspace{0.5cm} W1 Flux (mJy)}\hspace{0.5cm} &
  \colhead{\hspace{0.5cm} W2 Flux (mJy)}\hspace{0.5cm}
  }
\startdata
55285.7 & 2.848 $\pm$ 0.067 & 4.997 $\pm$ 0.162 \\
55477.4 & 2.962 $\pm$ 0.084 & 5.091 $\pm$ 0.130 \\
56747.8 & 3.169 $\pm$ 0.096 & 4.823 $\pm$ 0.154 \\
56941.7 & 3.002 $\pm$ 0.073 & 4.840 $\pm$ 0.136 \\
57108.2 & 3.161 $\pm$ 0.083 & 4.971 $\pm$ 0.146 \\
57302.7 & 3.083 $\pm$ 0.065 & 4.883 $\pm$ 0.148 \\
57468.7 & 3.149 $\pm$ 0.080 & 5.031 $\pm$ 0.135 \\
57668.5 & 3.120 $\pm$ 0.077 & 5.147 $\pm$ 0.145 \\
57829.7 & 3.233 $\pm$ 0.073 & 5.172 $\pm$ 0.153 \\
58034.7 & 3.520 $\pm$ 0.069 & 5.938 $\pm$ 0.141 \\
58189.9 & 4.624 $\pm$ 0.089 & 8.354 $\pm$ 0.135 \\
58399.6 & 4.225 $\pm$ 0.082 & 8.155 $\pm$ 0.149 \\
58555.7 & 4.418 $\pm$ 0.100 & 8.143 $\pm$ 0.147 \\
58763.9 & 3.814 $\pm$ 0.086 & 7.203 $\pm$ 0.139 \\
58921.2 & 3.869 $\pm$ 0.071 & 6.717 $\pm$ 0.141 \\
59129.5 & 3.417 $\pm$ 0.085 & 6.405 $\pm$ 0.145 \\
59285.4 & 3.595 $\pm$ 0.072 & 6.262 $\pm$ 0.144 \\
59495.2 & 3.268 $\pm$ 0.092 & 5.991 $\pm$ 0.156 \\
59651.1 & 3.425 $\pm$ 0.072 & 5.574 $\pm$ 0.156 \\
59859.3 & 3.317 $\pm$ 0.078 & 5.624 $\pm$ 0.173 \\
60016.7 & 3.401 $\pm$ 0.078 & 5.686 $\pm$ 0.136 \\
60225.0 & 3.217 $\pm$ 0.087 & 5.479 $\pm$ 0.146 \\
60383.6 & 3.361 $\pm$ 0.097 & 5.371 $\pm$ 0.146 \\
\enddata
\tablecomments{NEOWISE Band W1 (3.4 $\mu$m) and Band W2 (4.6 $\mu$m) photometry for SPIRITS 19q obtained from the unWISE image stacks as described in section \ref{sec:wise_photo} \citep{Lang:2014AJ....147..108L, Meisner:2017AJ....153...38M, Meisner:2017AJ....154..161M, Meisner:2023AJ....165...36M}.}
\end{deluxetable}

\subsection{Late-time MIR Photometry with \textit{Spitzer}/IRAC and NEOWISE}
\label{sec:wise_photo}
\begin{figure*}
    \centering
    \includegraphics[width=0.48\linewidth]{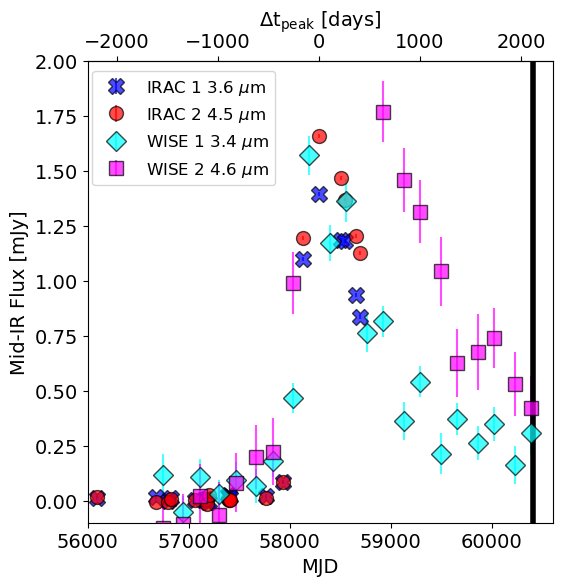} 
    \hfill
    \includegraphics[width=0.51\linewidth]{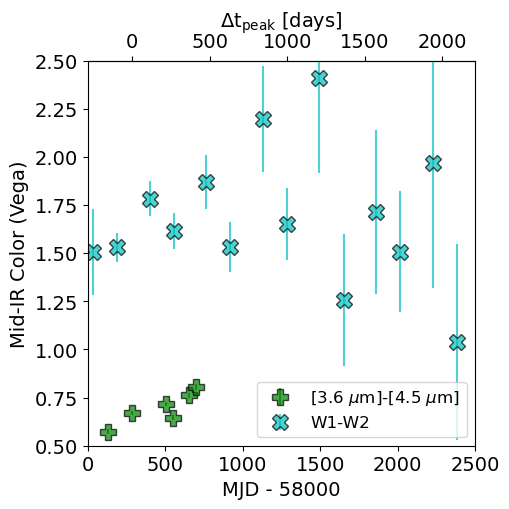} 
    \caption{\textit{Left:} Mid-IR baseline subtracted photometry for SPIRITS 19q from WISE/NEOWISE and \textit{Spitzer}. The total flux in the NIR has been decreasing since the peak of the outburst in 2019 suggesting that the newly formed dust is moving outward and cooling such that the peak of the emission is shifting to longer wavelengths. The \textit{JWST} NIRSpec IFU observations were obtained on UT 2024-04-03 (MJD 60403.91) denoted by a solid black line. \textit{Right:} Mid-IR color evolution in WISE W2-W1 and \textit{Spitzer}/IRAC [3.6 $\mu$m]-[4.5 $\mu$m]. The transient has been reddening since the initial outburst further suggesting that the newly formed dust is cooling, although at late times the uncertainty on the WISE color grows large.}
    \label{fig:lightcurve}
\end{figure*}

The location of SPIRITS 19q was covered by the NEOWISE \citep{Mainzer:2011ApJ...743..156M} all-sky mid-IR survey in W1 (3.4 $\mu$m) and W2 (4.6 $\mu$m). In Figure \ref{fig:lightcurve} and Table \ref{tab:photometry} we present W1 and W2 photometric data for the transient in forced photometry from the unWISE image stacks \citep{Lang:2014AJ....147..108L, Meisner:2017AJ....153...38M, Meisner:2017AJ....154..161M, Meisner:2023AJ....165...36M}, following the methods of \citep[][supplementary information 3]{De:2023Natur.617...55D}. In Figure \ref{fig:lightcurve} we also present \textit{Spitzer}/IRAC data in channel 1 (3.6 $\mu$m) and channel 2 (4.5 $\mu$m) from \citet{Lau:2021}. These data have had the quiescent flux of SPIRITS 19q subtracted off (IRAC channel 1 baseline flux = 0.34 mJy, IRAC channel 2 baseline flux = 0.32 mJy, WISE band 1 baseline flux = 3.05 mJy, WISE band 2 baseline flux = 4.95 mJy). The WISE baseline flux is higher than the \textit{Spitzer} baseline flux because of its lower spatial resolution ($\sim$ 6.5" for WISE versus $\sim$ 2" for \textit{Spitzer}/IRAC) which results in more flux from the bright \ion{H}{2} region where SPIRITS 19q is located being included in the source flux. 

At the time of our \textit{JWST} observations SPIRITS 19q, while significantly fainter than at peak, had not yet faded back to its quiescent level. Importantly, SPIRITS 19q is expected to be brighter at 4.6 $\mu$m than at 3.4 $\mu$m. The decreasing NIR luminosity suggests that the dust formed in the initial outburst has since cooled and the peak of the emission is now at longer wavelengths. The 3.6 $\mu$m $-$ 4.5 $\mu$m color at peak was already suggestive of a relatively cool (700 K) dust component which was chiefly emitting in the MIR \citep[][]{Lau:2021}

\section{Results and Analysis} 
\label{sec:Results and Analysis}

\begin{figure*}
    \centering
    \includegraphics[scale=0.7]{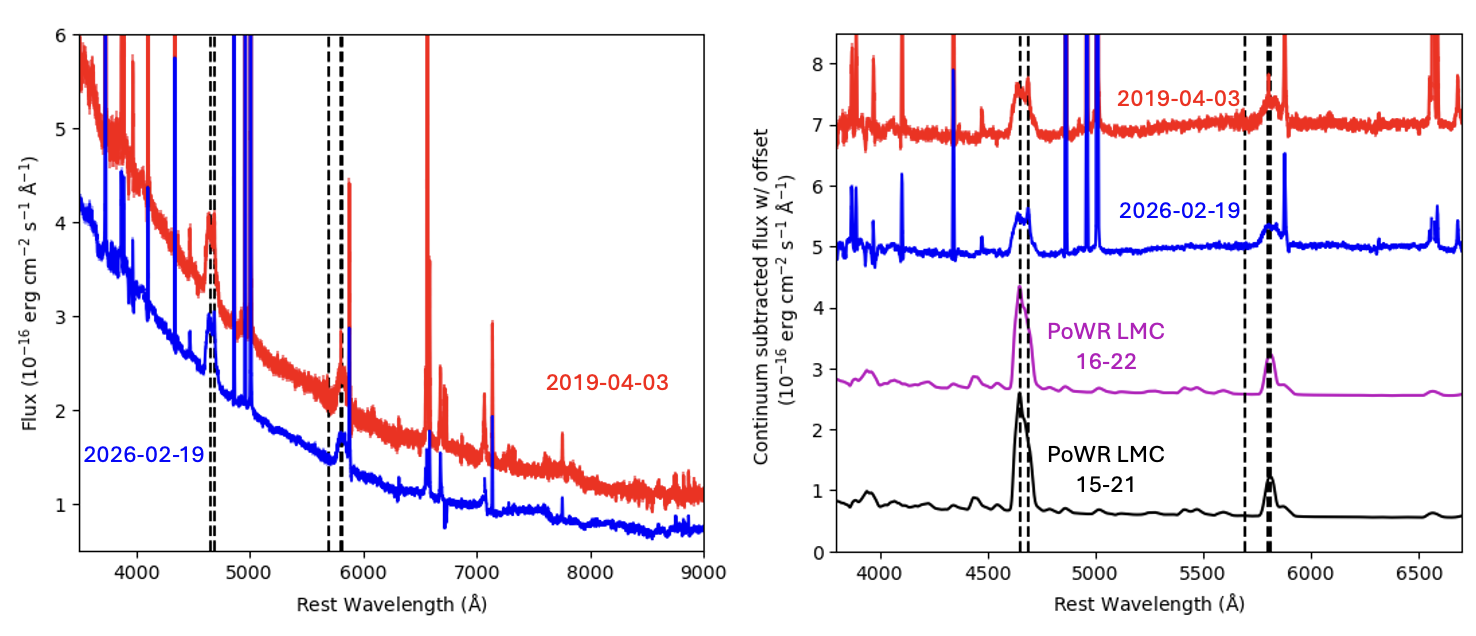}
    \caption{\textit{Left:} Reduced and flux-calibrated Keck I/LRIS spectra of SPIRITS 19q. The 2019-04-03 spectrum was originally presented in \citet{Lau:2021}. The locations of the C\,\textsc{iii-iv} $\lambda$4650, He\,\textsc{ii} $\lambda$4686, C\,\textsc{iii} $\lambda$5696, and C\,\textsc{iv} $\lambda$5801/5812 features are shown with dashed lines. The two spectra show nearly identical features suggesting that there was not strong contamination from the shock interaction region in the initial spectrum presented by \citet{Lau:2021}.  \textit{Right:} Keck I LRIS spectra of SPIRITS 19q with the continuum subtracted and an offset applied, as well as models which show similar features from the PoWR LMC WC grid smoothed to match the resolution of LRIS as described in section \ref{PoWR models}. The dashed lines mark the location of C\,\textsc{iii-iv}/He\,\textsc{ii} lines as described above. The absence of C\,\textsc{iii} $\lambda$5696 and the strength of the C\,\textsc{iv} $\lambda$5801/5812 doublet suggest an early-type WC star.}
    \label{fig:LRIS}
\end{figure*}

\begin{figure*}
    \centering
    \includegraphics[width=0.35\linewidth]{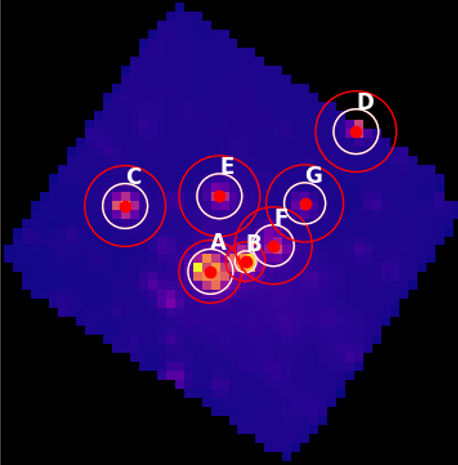} 
    \hfill
    \includegraphics[width=0.64\linewidth]{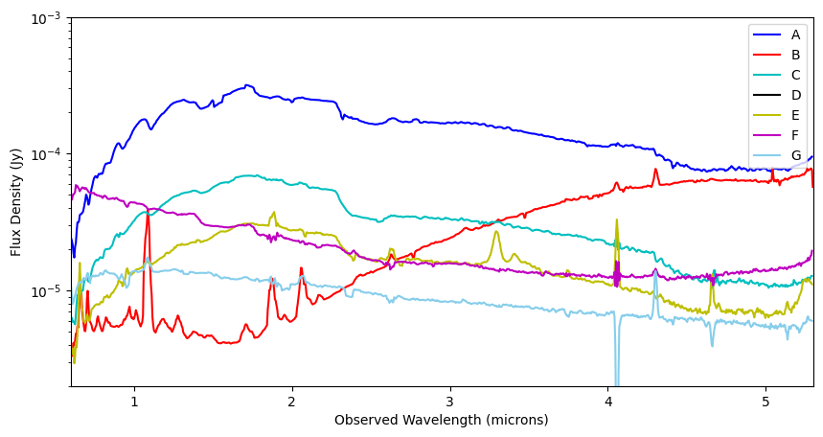} 
    \caption{\textit{Left:} IFU image of the region containing SPIRITS 19q integrated over the NIRSpec PRISM bandpass from 0.6 to 5.3 $\mu$m with 7 IR-bright sources identified. Circular aperture photometry (white circle) with a median background subtraction based on the cylindrical apertures shown in red is performed for each source in each wavelength slice. \textit{Right:} NIR spectra for each of the 7 sources identified in the IFU FOV. We identify source B as SPIRITS 19q as it is the only source which is brighter at 4.5 $\mu$m than it is at 3.6 $\mu$m. In addition to the red continuum of the dusty outburst, source B also shows WC features, confirming the association of the WC star with the dusty outburst.}
    \label{fig:identify}
\end{figure*}

\begin{figure}
    \centering
    \includegraphics[width=1.0\linewidth]{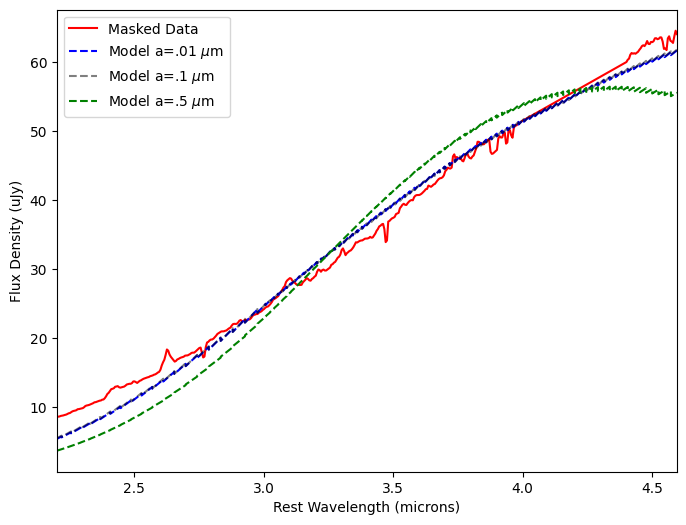}
    \caption{The red continuum of SPIRITS 19q with the spectral features masked. We fit to various size amorphous carbon dust grain models with optical constants from \citet{Zubko:1996MNRAS.282.1321Z}. The model favors smaller dust grains.}
    \label{fig:dust_model}
\end{figure}

\begin{figure}
    \centering
    \includegraphics[width=1.0\linewidth]{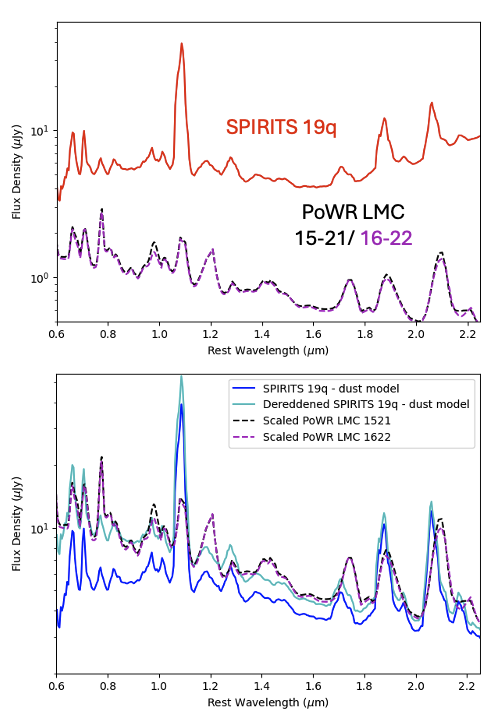}
    \caption{ \textit{Top:} The unscaled PoWR LMC WC 15-21 and 16-22 model stars spectra after correcting for instrumental broadening from 0.6 to 2.2 $\mu$m with the SPIRITS 19q spectrum. The models show the major features of the observed spectrum. \textit{Bottom:} The same PoWR LMC WC models scaled by 7.5 to match the continuum flux of the SPIRITS 19q spectrum with the dust continuum (based on the 0.1 $\mu$m grain size) subtracted off and a reddening correction applied. }
    \label{fig:mid_IR_PoWR}
\end{figure}

\subsection{Association of a Carbon-rich Wolf-Rayet Star with the Dusty Outburst}
\label{WC Association}

Photometry of SPIRITS 19q from \textit{Spitzer} and NEOWISE (see Figure \ref{fig:lightcurve}) identify it as a very red source (redF$^{4.6\mu m}_{\nu} >$ F$^{3.4\mu m}_{\nu}$) in the wavelength regime of our NIRSpec/IFU observations. In the spatially resolved IFU stack we identify 7 IR-bright sources within the bright crowded region which hosts SPIRITS 19q (see Figure \ref{fig:identify}). Extracting a spectrum for each of these 7 sources (Figure \ref{fig:identify}) using the circular apertures with cylindrical background regions shown, we identify source B as SPIRITS 19q as it is the only source which shows prominent carbon and helium emission lines as
well as the red continuum expected of the dusty transient. Creating a false color RGB image from the IFU data (blue 0.6--2.2 $\mu$m, green 2.2--3.7 $\mu$m, and red 3.7--5.3 $\mu$m) SPIRITS 19q (source B) is clearly the reddest object within the bright crowded region and has a spatial position consistent with the measured position of SPIRITS 19q from \textit{Spitzer}. In addition to the dusty continuum, the source B spectrum also shows the strong He 10830 $\AA$ feature previously observed by \citet{Lau:2021}, confirming the association of SPIRITS 19q with the WC star.

We note that \citet{Lau:2021} observed CO absorption in a ground based NIR spectrum of SPIRITS 19q, which would be unusual for a WC star which have C/O ratios greater than 1 \citep[e.g.,][]{Aadland:2022ApJ...924...44A}. In our resolved spectra from JWST/NIRSpec IFU it is clear that the source we identify as SPIRITS 19q does not show CO absorption. Sources A, C, and E do show CO absorption. The ground-based spectrum is likely contaminated by sources A, C, and E, resolving the conundrum. 

\subsection{Interpreting the WC Star Features with the Potsdam Wolf-Rayet (PoWR) Models}
\label{PoWR models}

From 0.6 to 2.3 $\mu$m the NIRSpec/IFU spectrum of SPIRITS 19q is dominated by features from the WC star. The relative strengths and shapes of these lines provide an important diagnostic tool to recover the physical parameters of the WC star responsible for the SPIRITS 19q outburst. We compare synthetic spectra generated from the Potsdam Wolf-Rayet (PoWR) models \citep{Grafner:2002A&A...387..244G, Hamann:2003A&A...410..993H, Sander:2015A&A...577A..13S} of WR atmospheres to map observed spectral features to stellar properties. 
We take advantage of the public model grid for WC stars \citep{Sander:2012A&A...540A.144S, Todt:2015A&A...579A..75T} at the metallicity of the LMC. The LMC WC models use X$_{\mathrm{Fe}}$ = 7 $\times$ 10$^{-4} (\approx$ 0.5 Z$_{\odot}$) which is close to measured metallicity of the \ion{H}{2} region VS~51 (0.6 Z$_{\odot}$) which hosts SPIRITS 19q. 

Because even the low resolution model spectra are higher resolution than our data we first apply a Gaussian filter to the model spectrum with a varying width as a function of wavelength based on the resolution of the NIRSpec PRISM disperser obtained from \textit{JWST} user documentation\footnote{\url{https://jwst-docs.stsci.edu/jwst-near-infrared-spectrograph/nirspec-instrumentation/nirspec-dispersers-and-filters\#gsc.tab=0}}, and resample the model at the observed wavelength for each IFU slice. We then scale the model flux to observed distance of NGC 2403 \citep[3.2 Mpc,][]{Radburn-Smith:2011} and directly compare to our observed spectrum of SPIRITS 19q from 0.6--2.3 $\mu$m. 

While the Keck I LRIS spectra are heavily contaminated by the surrounding HII region, they also show WC features. The LRIS data are higher resolution than the NIRSpec IFU data, so we take advantage of the high resolution line spectra models from the PoWR model grid. We apply a Gaussian filter to the model based on the resolution of LRIS as well as scaling to the distance of NGC 2403. The Keck I LRIS spectrum shows a strong C\,\textsc{iv} $\lambda$5801/5812, while the C\,\textsc{iii} $\lambda$5696 is undetected implying an early type WC star \citep{Crowther:1998MNRAS.296..367C}.

The PoWR model grid uses a 2-dimensional parameter space to organize models, T$_{*}$ and R$_{\mathrm{t}}$. T$_{*}$ is defined as the temperature at the radius where the Rosseland continuum optical depth is 20, as opposed to the usual definition of the effective temperature which is usually defined at a (total Rosseland) optical depth of 2/3. Because WR stars have such strong winds and high mass loss, T$_{2/3}$ in WR atmospheres is often not in the (quasi-)hydrostatic regime, but out in the wind.  T$_{*}$ is thus defined at a higher optical depth, aiming to better represent the hydrostatic layers \citep[though see, e.g.,][for remaining caveats]{Lefever:2023MNRAS.521.1374L,Lefever:2026A&A...707A...8L}. The transformed radius (R$_{\mathrm{t}}$) is a function of the WR wind velocity at infinity (v$_{\infty}$), the radius at which the Rosseland continuum optical depth is 20 (R$_{*}$), the mass loss rate ($\dot{\mathrm{M}}$), and a clumping factor (D) which describes the filling factor of optically thin clumps (which fill D$^{-1}$ of the volume and space between clumps is void): \begin{equation}
    R_{\mathrm{t}} = R_{*}\left( \frac{v_{\infty}}{2500\,\mathrm{km}\,\mathrm{s}^{-1}}/\frac{\dot{\mathrm{M}}\sqrt{\mathrm{D}}}{10^{-4}\, \mathrm{M}_{\odot}\,\mathrm{yr}^{-1}} \right)^{2/3}  \,
\end{equation}
The quantity $R_\mathrm{t}$ was introduced by \citet{Schmutz:1989A&A...210..236S} to account for the fact that for a variety of stellar temperatures the different parameter combinations leading to the same $R_\mathrm{t}$ yield may yield very similar spectra. 
For WRs with dense winds, the whole spectrum, including the continuum, is formed far out in the wind where the velocity is already a significant fraction of the terminal velocity \citep{Hamann:2004A&A...427..697H, Sander:2020MNRAS.491.4406S}. Therefore, as long as the mass-loss rates are similar, T$_{*}$ will not have a significant impact on the resulting spectrum. While a change in T$_{*}$ implies a change in the (assumed) hydrostatic radius, these deep layers are completely obscured by the dense wind (R$_{2/3} \gg $ R$_{*}$) and thus there is no change in the observed spectrum. High mass-loss models (implying a low $R_\mathrm{t}$) are thus relatively insensitive to T$_{*}$. 

We visually compare each of the 207 models in the LMC WC grid to the observed spectra of SPIRITS 19q. We find that models with very low $R_{\mathrm{t}}$ best resemble our data (and therefore high mass loss rates) for a variety of stellar temperatures. We find the most similar models to SPIRITS 19q to be the PoWR models LMC WC 15-21 and 16-22 (Figure \ref{fig:LRIS}, Figure \ref{fig:mid_IR_PoWR}).

While the LRIS spectrum is significantly contaminated by the flux of nearby objects, the JWST NIRSpec IFU spectrum is spatially resolved (Figure \ref{fig:spirits19q_inset_fig}). With improved constraints on the continuum flux, we can better compare the PoWR models to the observed SPIRITS 19q spectrum. We first take our best model for the dust continuum (see section \ref{Evidence of Dust}) and subtract it off. We then apply a reddening correction using the python package \texttt{dust\_extinction} \citep{Gordon:2024JOSS....9.7023G}. 
A variety of extinction models and R(V) terms are used in the literature for Wolf-Rayet stars. We use the \citet{Gordon:2023ApJ...950...86G} extinction model, implemented as G23 in the \texttt{dust\_extinction} code, with an R(V) = 3.1 and vary the magnitude of the extinction (parameterized by A(V)) to match the continuum slope of the PoWR models LMC WC 15-21 and 16-22. We also scale the models to match the continuum flux of the dereddened and dust continuum subtracted SPIRITS 19q spectrum. We find the best agreement between the models and observation for an A(V) of 1 and a PoWR model scale factor of 7.5 (Figure \ref{fig:mid_IR_PoWR}). 
This scale factor is uncertain due to contamination from other sources aside from the WC star in region B. At the very least there is contamination from the massive companion in the WC binary. Thus the values derived below should be treated as limits rather than precise estimates of the properties of the WC star in the SPIRITS 19q binary system.

After the work of \citet{Sander:2019A&A...621A..92S} we use this scale factor to obtain new measurements for the bolometric luminosity and mass loss rate of SPIRITS 19q. To preserve the model shape and line ratios $R_\mathrm{t}$ and temperature are fixed. The change is in the luminosity and $R_*$ which in turn changes the inferred mass loss rate. Using the Stefan-Boltzmann equation:

\begin{equation}
    L = 4\pi R_*^2\sigma_{SB}T_*^4
\end{equation}
as well as the equation for $R_\mathrm{t}$, and fixing the terminal wind velocity (v$_{\infty}$) and clumping factor (D),  it is possible to derive the scaling of $R_*$ and $\dot{\mathrm{M}}$ with luminosity as:
\begin{equation}
    \Delta \, \mathrm{log} \, R_* = 0.5 \cdot \Delta \, \mathrm{log} \, L
\end{equation}

\begin{equation}
    \Delta \, \mathrm{log} \, \dot{\mathrm{M}} = 0.75 \cdot \Delta \, \mathrm{log} \, L
\end{equation}

The parameters of these closest match scaled model stars for SPIRITS 19q are shown in table \ref{tab:model_WC}.

The comparison to the PoWR models suggest that the SPIRITS 19q WC star may have a higher than typical mass loss rate and may be more luminous as well. Galactic WC stars tend to have luminosities between 5 $\lesssim$ log $\frac{L}{L_{\odot}}$ $\lesssim$ 6 and mass loss rates between -5.3 $\lesssim$ log $\frac{\dot{M}}{M_{\odot} \, yr^-1}$ $\lesssim$ -4.2 \citep{Sander:2019A&A...621A..92S}. At lower metallicity WR stars are expected to be brighter as in order to produce a WR type spectrum it is necessary to be near the Eddington limit which is higher at lower metallicities \citep[e.g.,][]{Sander:2020MNRAS.491.4406S, Shenar:2020A&A...634A..79S}. For example the low metallicity SMC WO star AB8 has $\frac{L}{L_{\odot}}$ $\simeq$ 6.2 \citep{Shenar:2016A&A...591A..22S}. We also note that the luminosity of the WC star is likely overestimated. Given that SPIRITS 19q is producing dust, it must be at least a binary and we do not account for the companion luminosity or contamination from other unknown sources in region B. Contamination from other sources may also explain the unusual features (particularly the abnormally strong He\,\textsc{i} 10830 $\mathrm{\AA}$ line) observed in SPIRITS 19q compared to early type WC stars in the LMC \citep[e.g., the observations presented in ][]{Grafener:1998A&A...329..190G, Crowther:2002A&A...392..653C, Aadland:2022ApJ...924...44A}. The relatively high-mass loss rate is expected given that the observed brightness of the 2019 outburst suggests a large amount of dust formation. 

\begin{deluxetable}{ccc}
\tablecaption{PoWR Models
    \label{tab:model_WC}}
\tablehead{
  \colhead{\hspace{1.25cm}Parameter}\hspace{1cm} & 
  \colhead{\hspace{0.5cm}15-21}\hspace{1.0cm} &
  \colhead{\hspace{0.5cm}16-22}\hspace{1.0cm}
  }
\startdata
log(T$_{*}$) & 5.10 K  & 5.15 K \\
log(T$_{2/3}$) & 4.81 K  & 4.82 K \\
log(L$_{*}$) & 6.2 $L_{\odot}$ & 6.2 $L_{\odot}$\\
R$_{*}$ & 2.5 $R_{\odot}$ & 1.9 $R_{\odot}$\\
v$_{\infty}$ & 2,000 km s$^{-1}$ & 2,000 km s$^{-1}$ \\
log($\dot{\mathrm{M}}$) & -3.98 $M_{\odot}$ yr$^{-1}$ & -3.98 $M_{\odot}$ yr$^{-1}$ \\
\enddata
\tablecomments{Scaled model parameters for the scaled PoWR LMC WC models which most resembled the observed spectrum of SPIRITS 19q. Note that T$_{*}$ is defined as the temperature at the radius where the Rosseland continuum optical depth is 20, as opposed to the usual definition of the effective temperature (here T$_{2/3}$)  which is usually defined at a (total Rosseland) optical depth of 2/3. R$_{*}$ is calculated based on L$_{*}$ and T$_{*}$. Note that v$_{\infty}$ is a fixed parameter in the model grid.}
\end{deluxetable}

\subsection{Evidence of Dust} 
\label{Evidence of Dust}

Analysis of the \textit{Spitzer}/IRAC photometry to estimate the dust production rate of SPIRITS 19q depends on assumptions about the size and composition \citep[][]{Lau:2021} of the dust over the 0.98 year period where mid-IR luminosity is increasing the rate of dust formation is estimated to be between 10$^{-6}$ and 10$^{-5}$ $M_{\odot}$ yr$^{-1}$ (see table 4 of \citet{Lau:2021} for details).
Assuming optical constants for carbonaceous dust grains, as the circumstellar dust around WC stars is observed to be carbon-rich \citep[e.g.,][]{Cherchneff:2000A&A...357..572C}, it is possible to use the measured fluxes in these two filters to estimate the dust mass and temperature. Under the assumption that the mid-IR dust emission is optically thin, the total mass of dust can be expressed as \begin{equation}
\label{eq:dust}
    M_{\mathrm{dust}} = \frac{(4/3)a\rho_{\mathrm{b}}F_{\nu}d^2}{Q_{\mathrm{abs}}(\nu, a)B_{\nu}(T_{\mathrm{dust}})} \, ,
\end{equation}
where $a$ is the grain radius, $\rho_{\mathrm{b}}$ is the bulk density of the dust, $F_{\nu}$ is the observed flux, $d$ is the distance to SPIRITS 19q, $Q_{\mathrm{abs}}(\nu, a)$ is the dust emission efficiency, and $B_{\nu}$ is the Planck function. With data in only two filters and three unknowns ($a$, $M_{\mathrm{dust}}$, and $T_{\mathrm{dust}}$) it is necessary to assume a grain size ($a$) to make a measurement of dust mass and temperature. Using optical constants for amorphous carbon \citep{Zubko:1996MNRAS.282.1321Z} and graphite \citep{Draine:2001ApJ...551..807D, Li:2001ApJ...554..778L} as well as assuming a uniform grain size of 0.01 $\mu$m, 0.1 $\mu$m, or 0.5 $\mu$m \citep{Lau:2021} estimated the total dust mass at the peak of the IR outburst to be somewhere in the range of $\sim$ (2--22) $\times$ 10$^{-6}$ $M_{\odot}$ depending on the size and composition of the grains. From archival \textit{Spitzer} observations of NGC 2403 \citep{Lau:2021}, we place a lower limit on the binary period of a potential episodic WC dust producer of $\gtrsim$ 12 years. Almost all Galactic dust-forming WC binaries are observed to have orbital periods of less than $\sim$ 30 years \citep{Lau:2020}. An orbital period of 12 years corresponds to an average dust formation rate of up to $\lesssim$ (2--18) $\times$ 10$^{-7}$ $M_{\odot}$ yr$^{-1}$. 
 
With the JWST NIRSpec IFU spectrum we are able to place constraints on grain size and composition. 
We perform a least squares analytical fit of equation \ref{eq:dust} to the MIR spectrum, after masking out clear spectral features, for fixed grain sizes of $a$ = 0.01, 0.1, and 0.5 $\mu$m assuming optical constants for amorphous carbon from \citet{Zubko:1996MNRAS.282.1321Z}. We find the continuum flux from 2.5 to 5.0 $\mu$m is best fit by amorphous carbon dust grains with $a$ = 0.1 $\mu$m (or smaller). As expected there is no difference in the model for grain size 0.01 $\mu$m and 0.1 $\mu$m (Figure \ref{fig:dust_model}). Q$_{abs}$ is proportional to $a$ when $a$ $<$ $\lambda$/2$\pi$.

Since the outburst, the 3.4 $\mu$m and 4.6 $\mu$m luminosities of SPIRITS 19q have been decreasing and reddening suggesting that the dust formed in the outburst has been moving outwards and cooling (Figure \ref{fig:lightcurve}). The NIRSpec IFU data provides a strong constraint on the warm dust component, however there is evidence of an additional colder component from 5 to 5.3 $\mu$m. The total mass in this colder component is poorly constrained as the turnover in the continuum is outside the range of our NIR observations. Additional photometry at longer wavelengths would help to constrain the total dust mass of SPIRITS 19q at late times.

Taking the grain size and composition for the best fit model of the NIRSpec IFU spectrum, and assuming that the dust composition and size has not changed significantly since formation, the total dust mass estimated based on the peak of the IR outburst is 6.6 $\pm$ 0.4 $\times$ 10$^{-6}$ $M_{\odot}$ \citep{Lau:2021}. 
Assuming a binary period of at least 12 years this corresponds to a period averaged dust production rate of $\lesssim$ 5.5 $\times$ 10$^{-7}$ $M_{\odot}$ yr$^{-1}$. While uncertain, the period is unlikely to be dramatically longer. If it were, it would have been unlikely to observe the system in outburst with only a few years of monitoring. If the period is longer, that would imply that such systems are relatively more common, also increasing the estimate of their dust contribution. We also note that almost all Galactic dust-forming WC binaries are observed to have orbital periods of less than $\sim$ 30 years \citep{Lau:2020}, further suggesting that the orbital period of the SPIRITS 19Q system is likely not more than a factor of a few longer than our lower limit of 12 years.
 
To estimate the relative dust contribution of SPIRITS 19q to the galaxy NGC 2403, we estimate the dust production of AGB stars in NGC 2403. The total dust production rate of all AGB stars in the Large Magellanic Cloud is estimated to be 1.4 $\times$ 10$^{-5}$ $M_{\odot}$ yr$^{-1}$ \citep{Riebel:2012ApJ...753...71R}. 
Unlike the LMC, NGC 2403 is not a satellite. It is a late-type spiral galaxy, with a morphology similar to M33, on the outskirts of the M81 galaxy group \citep{Hunt:2025A&A...697A...9H}. NGC 2403 has mean global metallicity of 12 + log(O/H) $\sim$ 8.4 \citep[based on metallicity measurements of many HII regions made by][]{Rogers:2021ApJ...915...21R}. This is slightly higher than but comparable to the LMC average value of 12 + log(O/H) = 8.37 $\pm$ 0.03 \citep{Dominguez:2022MNRAS.517.4497D}.
NGC 2403 has a total stellar mass of 3.7 $\times$ 10$^{9}$ $M_{\odot}$ \citep{Leroy:2019ApJS..244...24L} which is $\sim$ 1.4 times the mass of the LMC \citep[ 2.7 $\times$ 10$^{9}$ $M_{\odot}$][]{vanderMarel:2006lgal.symp...47V}.
The current star formation rate of NGC 2403 is uncertain, with measurements between $\sim$ 0.5-1.0 $M_{\odot}$ yr$^{-1}$ \citep{Heald:2012A&A...544C...1H, Kennicutt:2009ApJ...703.1672K}. 
This is between 2-5 times higher than the LMC rate of $\sim$ 0.2 $M_{\odot}$ yr$^{-1}$ \citep{Harris:2009AJ....138.1243H}.

Under the assumptions that AGB dust production and star formation history are similar between NGC 2403 and the LMC we estimate that the total AGB dust production in NGC 2403 is $\sim$ 2 $\times$ 10$^{-5}$ $M_{\odot}$ yr$^{-1}$ by scaling from the LMC value using the ratio of total stellar mass. These are both major assumptions, such that this value should be treated approximately, and may only be accurate to within an order of magnitude. Given the relatively higher star formation rate of NGC 2403 this value is sensitive to differences in the star formation histories of the LMC and NGC 2403. Even with this uncertainty and the uncertainty in the rate of dust formation by SPIRITS 19q, that a single WC system can contribute dust at a rate as high as 3$\%$ that of tens of thousands of lower mass AGB stars is impressive. This suggests that such systems could be important dust contributors in the early Universe before AGB stars have time to evolve to form dust.

\section{Discussion} 
\label{sec:Discussion}
Dust grains---the seeds of star and planet formation---play a key role in regulating the physical and chemical conditions of the interstellar medium and are direct tracers of the chemical enrichment of the Universe. 
Despite the importance of dust in tracing our own cosmic origins, the origins of dust are unclear.
We know that low-mass stars can form significant quantities of dust on the asymptotic giant branch (AGB), and may even dominate dust production in the local universe \citep[e.g.,][]{Schneider:2024A&ARv..32....2S} but the slow evolution of such stars means that they likely cannot explain early-time dust production. Recent studies by JWST have shown that dust is formed in significant quantities even at very early times, within 600 million years of the Big Bang \citep{Witstok:2023}, suggesting the importance of dust formation by more massive stars. Rest frame UV spectroscopy of high redshift galaxies have also yielded detections of the 2175 $\mathrm{\AA}$ bump which is associated with carbonaceous dust grains \citep{Ormerod:2025MNRAS.542.1136O}. Supernovae, which have previously been invoked to explain the bulk of early universe dust formation, produce primarily dust grains with oxygen rich chemistry \citep{Gall:2014Natur.511..326G,Sarangi:2022A&A...668A..57S, Shahbandeh:2023MNRAS.523.6048S, Shahbandeh:2025ApJ...985..262S, Tinyanont:2025ApJ...985..198T}. It is also uncertain how much newly formed dust will survive the passage of the SN reverse shock 500--1,000 years post-explosion \citep[e.g.,][]{Micelotta:2016A&A...590A..65M}. 

By confirming the association of the highly dust-producing IR-outburst SPIRITS 19q with a WC binary system in the subsolar-metallicity \ion{H}{2} region of NGC 2403 (0.6 Z$_{\odot}$) we have shown that it is possible for WC systems to form at lower metallicity and contribute to the cosmic dust budget. The existence of the SPIRITS 19q system is an important piece of evidence which supports the conclusions of \citet{Lau:2020} which, by incorporating dust production into the Binary Population and Spectral Synthesis (BPASS) code of \citet{Eldridge:2017}, showed that WC systems can produce significant amounts of dust at lower metallicities despite being historically overlooked in the literature. 
Dust formed by WC stars, unlike the bulk of dust formed by supernovae, will be predominantly carbon-rich and may help to explain the large amounts of carbon-rich dust observed at high redshift before carbon-rich AGB stars can evolve \citep{Schneider:2024A&ARv..32....2S}.

Most known dust producing WC systems are found in the solar circle of our own galaxy, biasing us toward solar and super-solar metallicities \citep[e.g.,][]{Williams:2019MNRAS.488.1282W}. Such systems tend to also host late-type WC stars with lower wind velocities as compared to the early-type WC stars found in the LMC \citep{Sander:2019A&A...621A..92S}. Low-metallicity WC stars in the LMC likely formed as a result of binary interactions which resulted in the loss of the hydrogen envelope. That SPIRITS 19q is also host to an early-type WC star is in line with this hypothesis. The relatively higher mass-loss rates of early-type WC stars, including SPIRITS 19q, may lead to higher dust production rates. This is borne out by the relatively high dust production rate of SPIRITS 19q compared to other WC dust formers \citep{Lau:2021}. In order to provide the most robust estimates on the dust production rate of SPIRITS 19q, longer wavelength mid-IR imaging beyond 5 $\mu$m is needed to cover beyond the peak of dust continuum emission.
We also advocate for monitoring of SPIRITS 19q in the NIR/MIR to catch the next dusty outburst and measure a true orbital period for the system.
While challenging with respect to the limitations in observable features due to the distance, a dedicated follow-up study with a tailored atmosphere analysis might help to better constrain the properties of the WC star, including information about its surface carbon abundances, luminosity, wind velocity, and mass loss rate. 

Episodic highly dust producing WC systems will present themselves as luminous IR transients. To better constrain the contribution of such systems to the cosmic dust budget a larger sample of such objects is necessary. It is also unclear if, even though the binary stripping mechanism, it is possible to form a stripped WC star at even lower metallicities ($\lesssim$ 0.5 Z$_{\odot}$). A search of nearby very low metallicity dwarf galaxies for such stripped stars is motivated by our results. Next generation all-sky surveys which probe the NIR/MIR wavelength regime including existing data from WISE/NEOWISE, the ongoing SPHEREx survey, and the upcoming Roman and NeoSurveyor mission will allow for such searches of dust-producing WC stars at low-metallicity.

\section{Conclusions} 
\label{sec:Conclusions}

The \textit{JWST}/NIRSpec IFU observations of SPIRITS 19q confirm the association of the 2019 dusty outburst observed by \textit{Spitzer} with a WC star. By comparing the model spectra from the Potsdam Wolf-Rayet (PoWR) model stars at the metallicity of the LMC we find that the SPIRITS 19q WC star has a particularly high mass-loss rate and luminosity, typical of early type WC stars.  We estimate the total mass of dust produced in the outburst to be 6.6 $\pm$ 0.4 $\times$ 10$^{-6}$ $M_{\odot}$. This corresponds to a period averaged dust production rate of $\lesssim$ 5.5 $\times$ 10$^{-7}$ $M_{\odot}$ yr$^{-1}$. These observations demonstrate that:
\begin{itemize}
    \item Highly dust producing WC binary systems can be formed at subsolar (0.6 Z$_{\odot}$) metallicities.
    \item Even a single such system can form a significant amount of dust.
    \item IR transient searches for dusty outbursts will find dust-forming WC binary star systems. 
    \item WC stars may be a significant source of carbonaceous dust grains in the early universe. To confirm that such systems can form at even lower metallicities ($<$ 0.5 Z$_{\odot}$), we are motivated to search for such systems in very low-metallicity dwarf galaxies in the local universe. 
\end{itemize}

\section{Acknowledgments}
\label{Acknowledgments}

This work is based on observations made with the NASA/ESA/CSA James Webb Space Telescope. The data were obtained from the Mikulski Archive for Space Telescopes at the Space Telescope Science Institute, which is operated by the Association of Universities for Research in Astronomy, Inc., under NASA contract NAS 5-03127 for JWST. These observations are associated with program \#1863. T.O. acknowledges the support by JSPS KAKENHI grant No. JP24K07087. J.S.-B. acknowledges the support received by the DGAPA UNAM-PAPIIT IG101025 project. AACS is supported by the
Deutsche Forschungsgemeinschaft (DFG, German Research Foundation) in the
form of an Emmy Noether Research Group – Project-ID 445674056
(SA4064/1-1, PI Sander). AACS further acknowledges financial support by
the Federal Ministry for Economic Affairs and Energy (BMWE) via the
Deutsches Zentrum f{\"u}r Luft- und Raumfahrt (DLR) grant 50 OR 2509 (PI
Sander). This work was supported by the Federal Ministry of Education
and Research (BMBF) and the Ministry of Science of Baden-W{\"u}rttemberg
(MWK) as part of the Excellence Strategy of the Federal Government and
the States. This project was co-funded by the European Union (Project
101183150 - OCEANS).
M.F.C. acknowledges support by NASA under award number 80GSFC24M0006.
This research has made use of the Keck Observatory Archive (KOA), which is operated by the W. M. Keck Observatory and the NASA Exoplanet Science Institute (NExScI), under contract with the National Aeronautics and Space Administration.
The authors thank Greg Sloan for helpful discussions regarding the formation of carbonaceous dust,  and Elizabeth Tarantino for helpful discussions on AGB dust production at low metallicities. 
We also thank the anonymous referee. Their feedback has greatly improved the quality of this work.

%

\vspace{5mm}
\facilities{JWST(NIRSpec), Spitzer(IRAC), WISE/NEOWISE, Keck I/LRIS}


\software{Astropy \citep{Astropy:2013, Astropy:2018}, Matplotlib \citep{Hunter:2007}, NumPy \citep{Walt:2011}, SciPy \citep{Virtanen:2019}, dust$\_$extinction \citep{Gordon:2024JOSS....9.7023G}
}

\bibliography{main}{}
\bibliographystyle{aasjournal}

\end{document}